\documentclass{PoS}
\usepackage{bm}

\def\Eq#1{Eq.~(\ref{#1})}

\def\<{\langle}
\def\>{\rangle}

\title{String tensions of $SU(N)$ gauge theories in 
$2+1$ dimensions}

\ShortTitle{String tensions of $SU(N)$ gauge 
theories in $2+1$ dimensions}

\author{\speaker{Barak Bringoltz} and Michael Teper\\
    University of Oxford\\
        E-mail: \email{barak@thphys.ox.ac.uk},
                                        \email{m.teper1@physics.ox.ac.uk}}

\abstract{We calculate the energy spectrum of closed strings in $SU(N)$ gauge theories with $N=2,3,4,6,8$ in $2+1$ dimensions to a high
  accuracy. We attempt to control all systematic errors, and this
  allows us to perform a precise comparison with different theoretical predictions. 

When we study the dependence of the string mass on its length $L$ we find that the Nambu-Goto
  prediction is a very good approximation down to relatively short
  lengths, where the L\"uscher term alone is insufficient. We then
  isolate the corrections to the L\"uscher term, and compare them to
recent theoretical predictions, which indeed seem to be mildly preferred by
the data.

When we take these corrections into account and extract string
tensions from the string masses, we find that their continuum limit is lower by $2\%-1\%$ from the predictions of
Karabli, Kim, and Nair. The discrepancy decreases with $N$, but when we
extrapolate our results to $N=\infty$ we still find a discrepancy of
$0.88\%$ which is a $4.5 \sigma$ effect.}

\FullConference{XXIVth International Symposium on Lattice Field Theory\\
                 July 23-28, 2006\\
                 Tucson, Arizona}

\begin{document}

\section{Introduction}

There are many indications for a connection between QCD and
 string theory, which appear in a wide range of
 experimental phenomena and 
theoretical works \cite{Polchinski}.
 Here we are interested on the QCD side and 
study the energy spectrum of the `original' 
string - the flux
tube of pure Yang-Mills - with an emphasis on its large-$N$ limit. We
focus on three Euclidean dimensions, and our motivation is
two-fold. First, we wish to perform a precise test of the remarkable work of Karabali, Kim,
and Nair \cite{KN}. This work yields the following
prediction for the string tension $\sigma$
\begin{equation}
\frac{\sqrt{\sigma}}{g^2N}=\sqrt{\frac{1-1/N^2}{8\pi}}, \label{sigma_KN}
\end{equation}
where $g$ is the
super-renormalised coupling\footnote{Recall that in
 $2+1$ dimensions the coupling $g^2$ has dimensions of mass.}.
A previous comparison of \Eq{sigma_KN} to lattice results for $N=2,3,4,6$
\cite{Teper_Lucini} showed that \Eq{sigma_KN} is higher from the data by about
$2\%-1\%$, but also that
this discrepancy decreases with $N$. This and the fact that 
 \cite{KN} predicts no
screening of zero $N$-ality charges suggests the possibility that 
the analysis in \cite{KN} may be exact at $N=\infty$.
This would appear to be contradicted by the $N\to \infty$ extrapolation of \cite{Teper_Lucini}, 
but the results presented there included several unestimated
systematic errors which, while small, may be significant at the $1\%$
level. Two of these lead to an underestimate of the lattice 
string tensions, and we check whether by controlling them, the lattice results become consistent with \Eq{sigma_KN} at $N=\infty$.

 The first error rises in the process of 
extracting the 
string tension from the string mass, where it
 is typical to neglect corrections that are sub-leading to the L\"uscher term.  Indeed, as the 
string length $L$ becomes larger, these corrections decrease as
$1/L^4$, but we wish to estimate them on a 
quantitative level. The way the string mass changes with $L$ is also interesting
theoretically (and this constitutes our second
 motivation for this study). It gives us
information on the effective string theory of the flux
tube, that can be compared with older \cite{old_works} 
and more recent \cite{new_works} predictions. The second systematic
error that we remove is the neglect
of $O(g^4)$ terms in the continuum extrapolation of the string tensions.

Here we present an analysis of these systematic errors, and in the case of the $O(1/L^4)$ error, 
we compare to the predictions of \cite{new_works}. 
After removing both
systematics we compare the resulting string
tensions to \Eq{sigma_KN}. The results presented here were obtained in a preliminary analysis of the data. Publications with a more
  extensive analysis, that will also include the raw data, are forthcoming \cite{papers_2_come}.

\section{Methodology}

There exists a plethora of lattice works studying properties of
flux-tubes by looking at open or closed strings (for example see the review
 in \cite{otherworks}).
 To avoid `contamination' from perturbative effects, and to compare
 directly with the spectrum of the effective
string theories, we choose to measure the large distance exponential fall-off of correlation functions of
strings closed around the spatial torus.
Our analysis proceeds in two
stages.

\subsection{Stage I - the string mass dependence on the string length}
 We first work with a fixed lattice spacing (by fixing the lattice
coupling) and study the dependence of the string mass $m$ on its
length $L$. Fitting these results we
are able to test
the theoretical predictions in
\cite{old_works,new_works}. In addition, these fits
provide us with a practical tool to extract string tensions from
strings which are shorter then usual,
and with a quantitative estimate of the systematic error induced
by using the L\"uscher term alone.

\subsection{Stage II - continuum extrapolation of string tensions}

Here we perform measurements of
string masses with a fixed physical length $L$ and different
lattice spacings. We choose $L\sqrt{\sigma}\stackrel{>}{_\sim} 3-3.5$ and use the fits we obtain
from stage I to extract string tensions. We then extrapolate these to the continuum limit and compare with
\Eq{sigma_KN} for all values of $N$ as well as for the
extrapolation to $N=\infty$.

\subsection{Lattice construction}
We define the gauge theory on a discretized periodic Euclidean three
dimensional space-time with $N_0\times N_1\times N_2$ sites, and perform Monte-Carlo simulations of a simple Wilson action.
We use the Kennedy-Pendelton heat bath algorithm for the link 
updates, followed by five over-relaxations of all the $SU(2)$ 
subgroups of $SU(N)$. We measure correlation functions of Polyskov loops that wind around the $\hat{0}$
direction so $L=aN_0$. 
The correlations are measured along direction
$\hat{1}$ and we project to zero transverse
 momentum by averaging over direction $\hat{2}$.
 Although here we mainly 
present results for the ground state, we study all
possible values of the $N$-ality $k=1,2,\dots,[N/2]$, and their first few excited states. 
Using improved operators \cite{Teper_et_al} we
are able to obtain overlaps which are 
almost perfect for the ground
state of $k=1$, but somewhat lower for the
 excited states. To avoid finite volume effects we
increase the lattice in the $\hat{1}$ and $\hat{2}$ directions
for the shorter strings \cite{Teper_et_al}. 
Stage-I is studied for $N=3,4,6,8$, and
$1.3-1.6\stackrel{<}{_\sim}L\sqrt{\sigma}\stackrel{<}{_\sim}3-6.2$, while
in stage-II we study $SU(2)$ as well and restrict to 
$L\sqrt{\sigma}\stackrel{>}{_\sim}3-3.5$ with $0.1 \stackrel{<}{_\sim}
a\sqrt{\sigma} \stackrel{<}{_\sim} 0.75$. The raw data will be presented in \cite{papers_2_come}. 

\section{Results - dependence of the string mass on its length}
\label{Ldep}

\begin{figure}[htb]
\centerline{
\includegraphics[width=7cm]{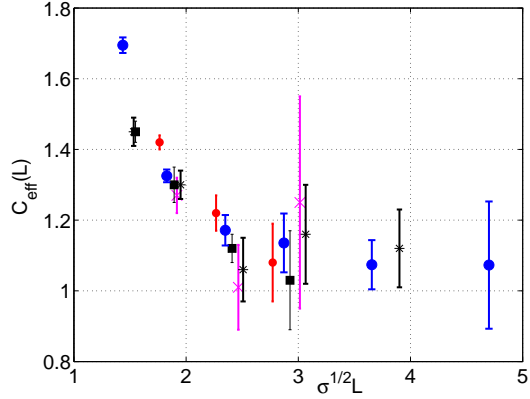} 
}
\caption{Effective central charge as a 
function of the string
  length, for $SU(3), \beta=14.7172$ (blue circles), $SU(4), \beta=28.00$
  (red dots), $SU(6), \beta=59.40, 90.00$ (black stars and squares), and $SU(8), \beta=108.00$
  (pink crosses).}
\label{fig1}
\end{figure}

 We first check the universality class of the string by looking at the behaviour of an effective
central charge defined as 
$\frac{m(L)}{L}\equiv \sigma  - \frac{\pi \, C_{\rm
    eff}(L)}{6L}$. Here 
$C_{\rm eff}(L)$ should be $1$ at $L\gg 1/\sqrt{\sigma}$ for 
the bosonic string, and is obtained
from our data by performing fits to successive pairs of adjacent points. The
results are shown in Fig.~\ref{fig1} where we see that for
$\sqrt{\sigma} L \stackrel{>}{_\sim} 3$ the charge $C_{\rm eff}$ becomes
consistent with $1$. 

\begin{figure}[htb]
\centerline{
\includegraphics[width=7cm]{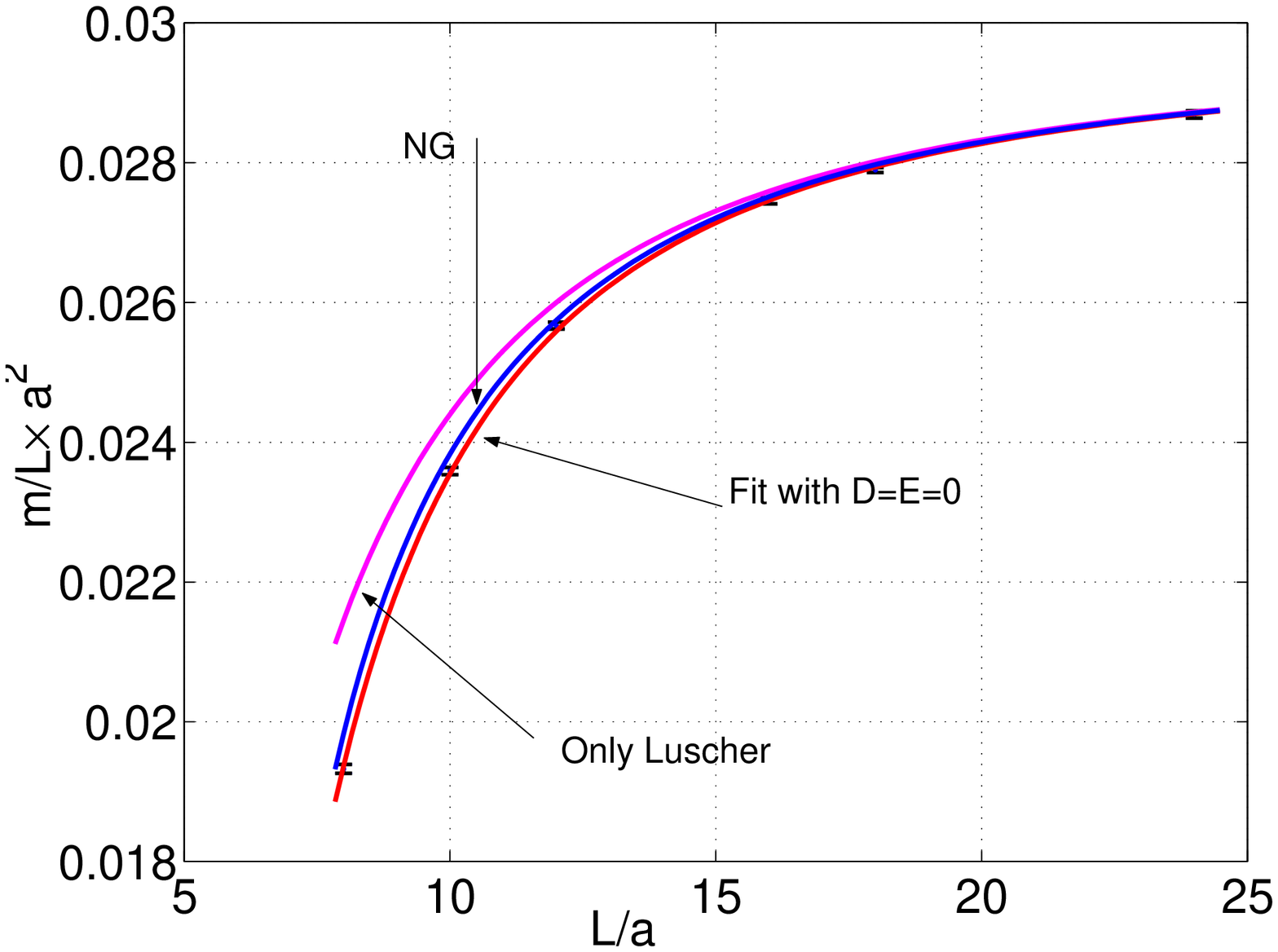} 
\qquad
\includegraphics[width=7cm]{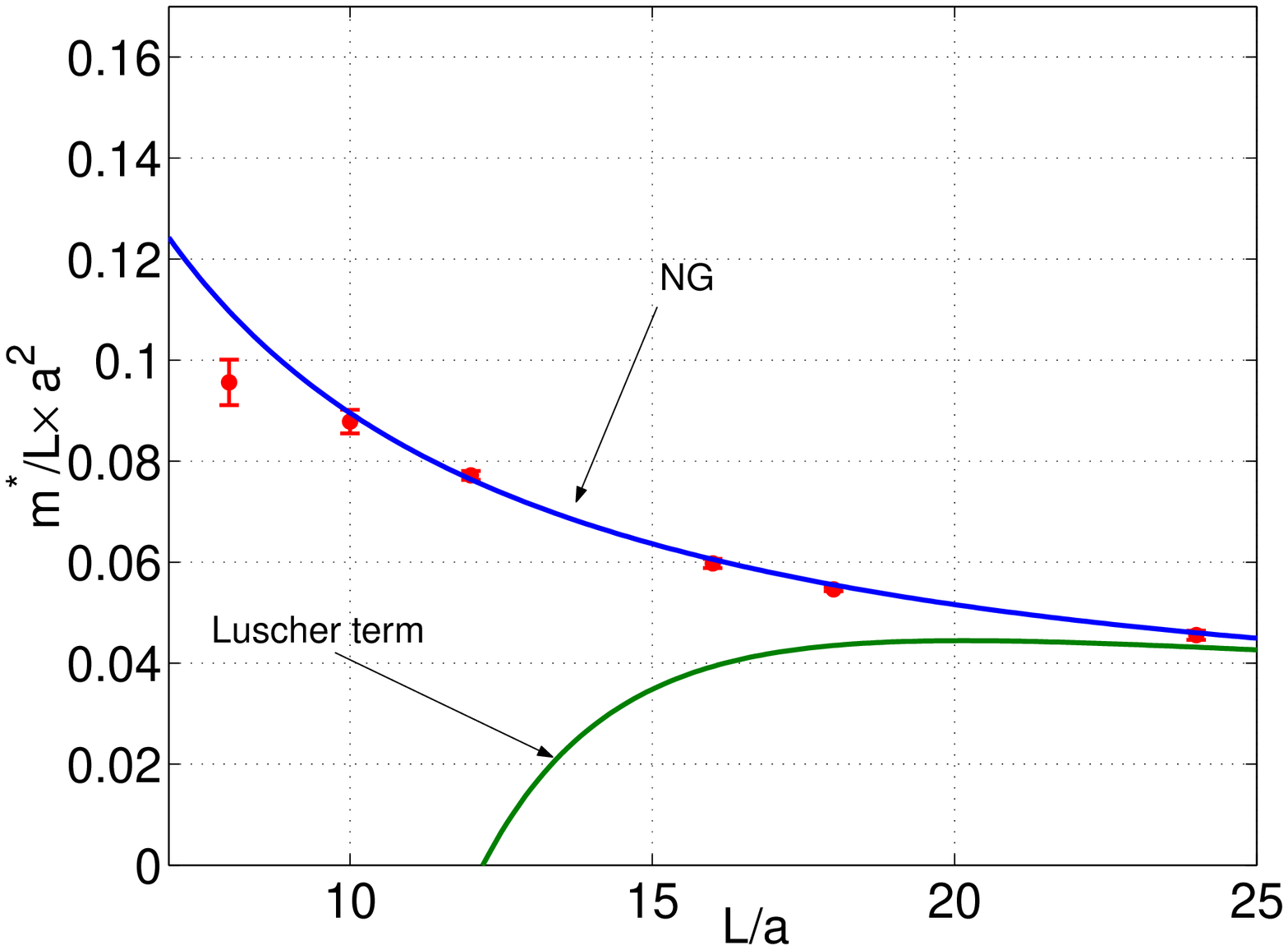}
}
\caption{\underbar{Left:} $m(L)/L$ for the ground state closed string of
  $SU(6)$ at $\beta=90.00$. The fit (red line) gives
  $a\sqrt{\sigma}=0.172148(70)$. Substituting this in $m_{\rm Luscher}$ and
  $m_{\rm NG}$ gives the magenta and blue lines. \underbar{Right:} $m(L)/L$ for the first  excited state of the $k=1$ string. The lines are plots of $m_{\rm
    Luscher}$ and $m_{\rm NG}$ for a quantum $n=1$ and  $\sigma$ as 
 extracted from the ground state.}
\label{fig2}
\end{figure}

Next we fit our data to the general form\footnote{The  $1/L^3$ term is missing since it was unacceptable fit for our data, and was suggested by L\"uscher and Wiesz to be disfavoured theoretically \cite{new_works}.} 
\begin{equation}
\frac{m(L)}{L} = \sigma - \frac{B}{L^2}-
\frac{C}{(\sigma L^2)L^2} - 
\frac{D}{(\sigma L^2)^{3/2}L^2} -
\frac{E}{(\sigma L^2)^2L^2}. \label{old_fit}
\end{equation}
Here we focus on the following fits. First we fix the $O(1/L^2)$ to
be the L\"uscher term with $B=\pi/6$ and (i) let $C$ be a free
fit parameter but fix $D=E=0$ (ii) follow the theoretical predictions
\cite{new_works} and constrain $C=\pi^2/72$ while fitting with $D\neq0,E=0$ or $D=0,E\neq0$.
We also compare our fits to what one obtains when one uses the L\"uscher
term alone
$m_{\rm{\,\,Luscher}}(L)/L = \sigma  - \frac{\pi}{6L^2}$, or the Nambu-Goto formula 
$m_{\rm NG}(L)/L = \sigma  \sqrt{1-\frac{\pi}{3\sigma
    L^2}}$ as obtained by Arvis \cite{old_works}. (Here $\sigma$ is the
string tension obtained from our fit)
 As a demonstration, we present results for $SU(6)$ and $\beta=90.00$ in the left panel
of Fig.~\ref{fig2}, where one can see that the L\"uscher term is a good
approximation for $\sqrt{\sigma} L\stackrel{>}{_\sim}3$ while the Nambu-Goto prediction works
remarkably well down to almost the deconfinement length, but not exactly. 

The results for all the other gauge groups are presented in Table~\ref{table1},
where one can see that the scaling and $1/N$ effects are relatively
small. Our best fits in the
case (i) and (ii) have comparable confidence levels of $43\%-90\%$, but a mild
preference is seen toward option (ii). In apparent contrast 
to that, 
the coefficient $C$ obtained in the fits of type (i) turns out to be about twice as
large as predicted in \cite{new_works}. Nonetheless, when 
we perform fits in which both $C$ and either $E$ are free
(denoted as type (iii) in the last
two columns in Table~\ref{table1})
 we find that the data is almost consistent with $C\simeq \pi^2/72$ as \cite{new_works} predicts.
\begin{table}[htb]
\centerline{
\begin{tabular}{|c|c|cc|cc|cc|}
\hline \hline
 \multicolumn{2}{|c|}{Data set}  & \multicolumn{2}{c|}{fit (i), \,$D=E=0$} &
\multicolumn{2}{c|}{fit (ii),\, $C=\pi^2/72$} & \multicolumn{2}{c|}{fit (iii), \,$D=0$}\\ \hline
$N$ & $\beta$& $a\sqrt{\sigma}$  & $\frac{C}{\pi^2/72}$ & $D$ & $E$  &
$\frac{C}{\pi^2/72}$ & $E$ \\ \hline
3 & 14.7172 & 0.261157(82) & 1.96(8) & 0.230(6) & 0.299(8) & 1.32(12)
& 0.23(3)\\ \hline
4 & 28.00 & 0.25198(16) & 2.26(8) & 0.23(4) & 0.36(2) & 1.36(32) & 0.25(9)\\ \hline
6 & 59.40 & 0.27870(12) & 1.91(6) & 0.17(1) & 0.25(2) & 1.45(23) & 0.12(6)\\ 
6 & 90.00 & 0.172148(70) & 1.85(6) & 0.16(1) & 0.22(2) & 1.21(24) &
0.16(6) \\ \hline
8 & 108.00 & 0.27402(34) & 1.58(2) & 0.13(4) & 0.19(6) & 0.77(32) & 0.28(9)\\
\hline \hline
\end{tabular}
}
\caption{The results of our fits of type i-iii (see text). The values
  of $a\sqrt{\sigma}$ for fits of type ii,iii were consistent mostly
  within $1.5\sigma$ with those for type i, which are presented in the
  table.
\label{table1}}
\end{table}

Finally we can use the value of
$\sigma$ obtained from the ground state of the string, and substitute
it in the L\"uscher and Nambu-Goto formulas, but with a
quantum $n=1$ (see for example the work of Arvis \cite{old_works}). We compare the result to the mass of the first excited
state that we measure. The comparison is presented in the right panel
of Fig.~\ref{fig2}. We see that the L\"uscher formula does not describe
the data at all, but that, surprisingly, the full
Nambu-Goto describes the data quite well.
 This seems to suggest that it is more natural to fit the data for
 $m^2$ rather than for $m$, since the former has the full
Nambu-Goto expression as a zeroth approximation. An analysis in this form will be presented in a forthcoming publication \cite{papers_2_come}.

To conclude this section we find that the systematic error induced by assuming $C=D=E=0$ for strings with 
$\sqrt{\sigma} L \simeq 3$ is to underestimate the string 
mass by about a third of a percent. In the next section we use this result and
extract string tensions using the form \Eq{old_fit} with
$B=\pi/6$, $D=E=0$ and $C$ as given by the fits to the different
gauge groups (we assume that the scaling
violations in $C$ are small, as observed for $SU(6)$ - see Table~\ref{table1}).

\section{Results - extrapolation of string tensions to the continuum}

\begin{figure}[htb]
\centerline{
\includegraphics[width=7cm]{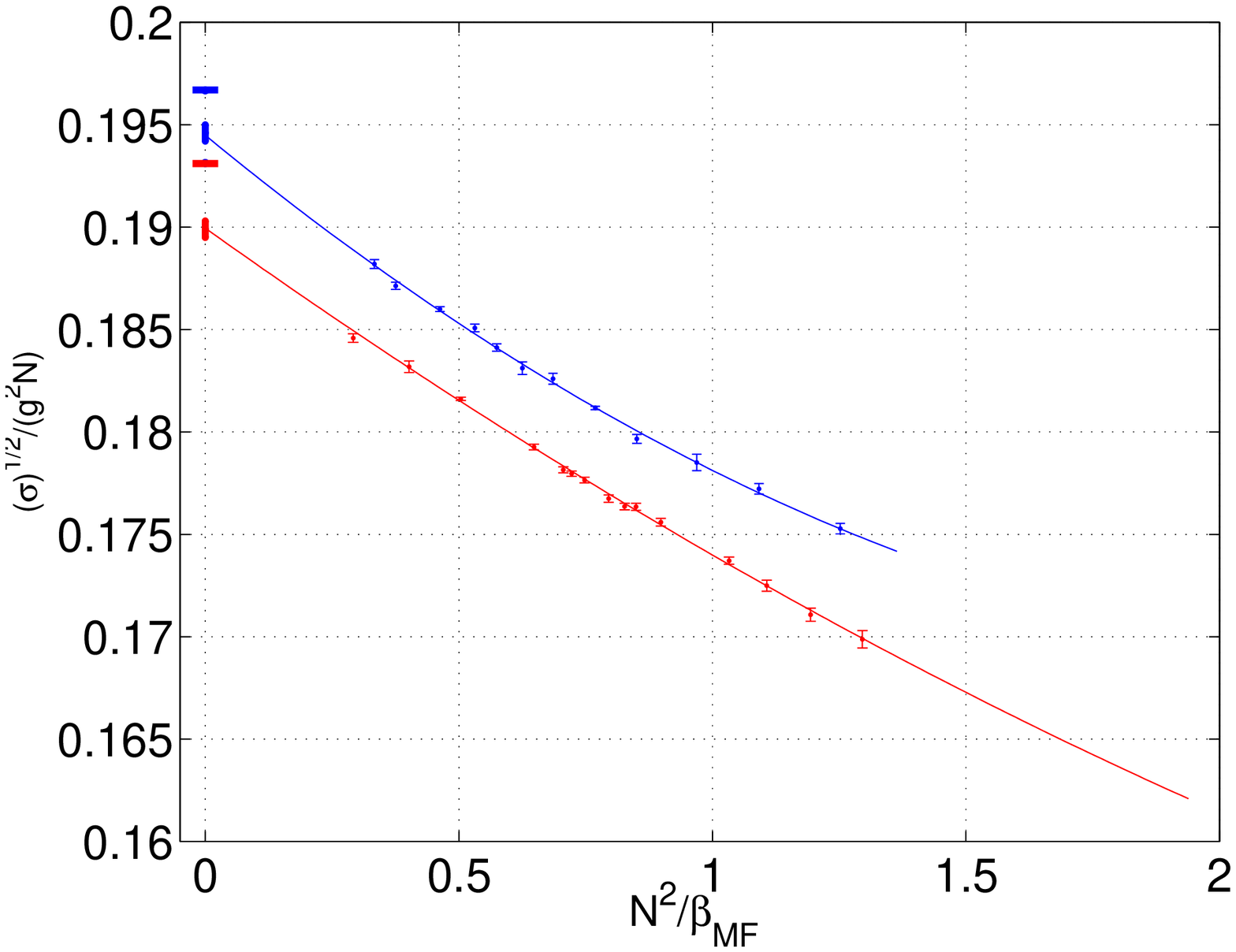} 
\qquad
\includegraphics[width=7cm]{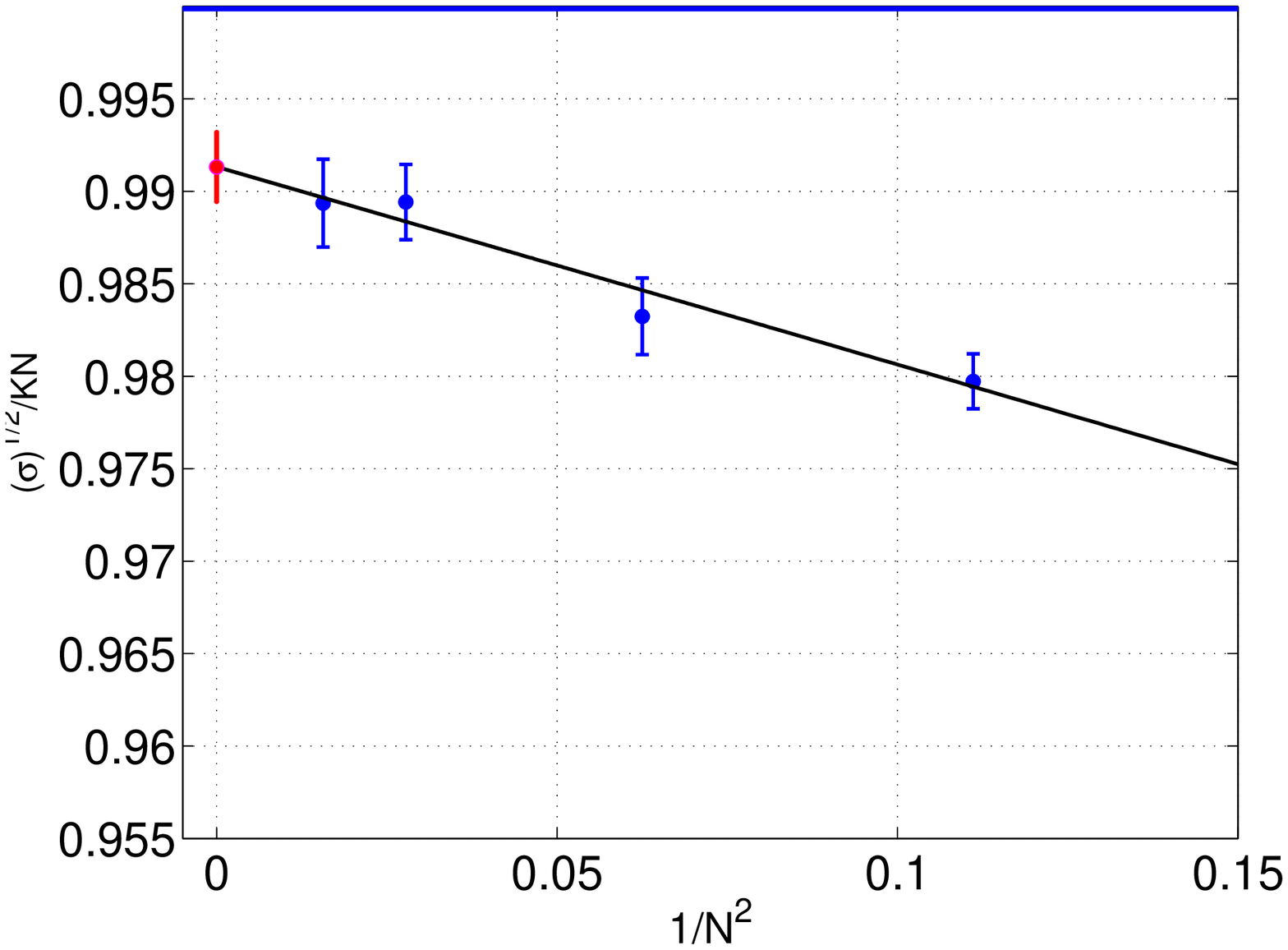} 
}
\caption{\underbar{Left:} The dimensionless quantity 
$\beta_{MF} \frac{a\sqrt{\sigma}}{2N^2}$ as a function of the improved
  coupling $1/\beta_{MF}$ for $SU(4)$ (lower plot, in red)  and
  $SU(6)$ (higher plot, in blue). The error bars at $1/\beta_{MF}=0$ denote the result of the
  continuum extrapolation, while the horizontal bars denote the values
  predicted by Karabali, Kim, and Nair \cite{KN}. \underbar{Right:} The ratio between the prediction of \cite{KN} and our data,
  $r$, as a function of $1/N^2$, for $N=3,4,6,8$. The error bar at $1/N^2=0$ denotes
  the extrapolation to the $N=\infty$ limit, while the horizontal line
  at $r=1$ denotes where the prediction of \cite{KN}
  exactly matches the lattice result.}
\label{fig3}
\end{figure}

Equipped with the fitting formula for $m(L)$ we extract string
tensions from a series of mass
 measurements where we keep the physical
length of the string to be $L\sqrt{\sigma}\stackrel{>}{_\sim}3-3.5$ and change the lattice
spacing by changing $\beta$. We perform a continuum extrapolation
with the ansatz
\begin{equation}
\beta_{MF}\frac{a\sqrt{\sigma}}{2N^2}=\left(\frac{\sqrt{\sigma}}{g^2N} \right)_{\rm continuum} - \frac{b_0}{\beta_{MF}} - \frac{b_1}{\beta^2_{MF}}, \label{continuum}
\end{equation}
Here, $\beta_{MF}$ is the mean field improved coupling
$\beta_{MF}=\beta \cdot\<u_p\>$, for which we measure the expectation value of
the lattice plaquettes $\<u_p\>$ at each
 $\beta$. Fitting with \Eq{continuum} 
is the way we remove the systematic error 
related to the $O(g^4)$ terms that we mention in the introduction. As a demonstration we
present the results for $SU(4)$ and $SU(6)$ in the left panel of
Fig.~\ref{fig3} where it can seen that $b_1>0$ and
therefore that the removal of this systematic error increases the
estimate of the continuum string tension.
Nonetheless, as seen
in the Figure, the extrapolation of the
string tension are still lower compared to the value predicted
by Karabali, Kim, and Nair (KKN) for these gauge
 groups. In the
right panel of Fig.~\ref{fig3}
 we present the ratio $r\equiv \frac{\left( \sqrt{\sigma}/g^2N\right)_{\rm
    KKN}}{\left(\sqrt{\sigma}/g^2N\right)_{\rm Lattice}}$ between
the Karabai-Kim-Nair prediction \cite{KN} and the lattice data (the numerical values are given in Table~\ref{table2}).

\section{Summary}

We measure string masses in $SU(N)$ gauge theories with $N=2,3,4,6,8$
in $2+1$ dimensions using lattice techniques. Here we present the
results of two studies performed with these masses. In the first we investigate the behaviour of
the closed string masses as a function of their length at a fixed
lattice spacing. We find that for $\sqrt{\sigma}L\stackrel{>}{_\sim}3$ our data is consistent with a L\"uscher
term of a unit central charge. For shorter strings the
L\"uscher term is insufficient and we find that the Nambu-Goto prediction
works surprisingly well, but is not exact. We fit the deviations from the L\"uscher term 
and compare the result to the recent
theoretical predictions \cite{new_works}, which seem to
 be mildly supported by the data.
We estimate that the systematic error induced by neglecting these 
sub-leading terms at $L\sqrt{\sigma}\simeq 3$ is to underestimate the string masses by about a
third of a percent.

In the second study we calculate string tensions by restricting to 
 string lengths of $\sqrt{\sigma}L \stackrel{>}{_\sim} 3-3.5$ 
and change the lattice spacing. Using the fit from the first study we
 extract the string tensions and  extrapolate to the continuum limit.
The results are
given in Table~\ref{table2}, where we also give the 
Karabali-Kim-Nair (KKN) predictions. As indicated in the table, when we
extrapolate to $N=\infty$ we find 
that the lattice result is lower by
about $0.88\%$ then the prediction of \cite{KN} which is a $4.5\sigma$
effect.

\begin{table}[htb]
\centering{
\begin{tabular}{|c|c|c|c|c|c|c|} \hline \hline
& $N=2$ & $N=3$ & $N=4$ & $N=6$ &  $N=8$ & $N=\infty$\\ \hline
KKN& $0.17275$ & $0.1881$ & $0.1931$ & $0.1967$ &  $0.19791$ & $0.19947$\\
\hline
Lattice & $0.16678(42)$ & $0.18425(28)$ & $0.1898(4)$ & $0.1946(4)$ &
$0.19580(47) $ & $0.1977(4)$\\ \hline \hline
\end{tabular}
}
\caption{$k=1$ string tensions for $SU(N)$ pure gauge theories in the continuum and the predictions of
  Karabali, Kim, and Nair (KKN) \cite{KN}.}
\label{table2}
\end{table}

\section{Future prospects}

An additional systematic
error is related to the contamination from
excited states, which we take partially into account by fitting correlation functions with a single
 exponential at large $t$. Eliminating this effect will tend to give a
 lower string tension, which would {\em increase} the discrepancy with
\Eq{sigma_KN}. Also, since in the case of $k$-strings our operators
have typically a lower overlap than in the case of $k=1$, it is
possible that one will obtain a significantly lower $k$-string
tension there. This may bring the results numerically much closer to Casimir
scaling, which may have interesting theoretical implications, and an initial
analysis of this issue will be present in \cite{papers_2_come}.
Other directions of research include a detailed spectrum calculation
of excited string states and of glueballs
\cite{work_with_andreas} with a larger
basis of operators that could
 be compared with recent theoretical predictions \cite{Minic_et_al}

\end{document}